# Definition and Treatment of Systematic Uncertainties in High Energy Physics and Astrophysics

Pekka K. Sinervo
*Department of Physics, University of Toronto, Toronto, ON M5S 1A7, CANADA*

Systematic uncertainties in high energy physics and astrophysics are often significant contributions to the overall uncertainty in a measurement, in many cases being comparable to the statistical uncertainties. However, consistent definition and practice is elusive, as there are few formal definitions and there exists significant ambiguity in what is defined as a systematic and statistical uncertainty in a given analysis. I will describe current practice, and recommend a definition and classification of systematic uncertainties that allows one to treat these sources of uncertainty in a consistent and robust fashion. Classical and Bayesian approaches will be contrasted.

## 1. Introduction to Systematic Uncertainties

Most measurements of physical quantities in high energy physics and astrophysics involve both a statistical uncertainty and an additional "systematic" uncertainty. Sytematic uncertainties play a key role in the measurement of physical quantities, as they are often of comparable scale to the statistical uncertainties. However, as I will illustrate, the definition of these two sources of uncertainty in a measurement is in practice not clearly defined, which leads to confusion and in some cases incorrect inferences. A coherent approach to systematic uncertainties is, however, possible and I will attempt to outline a framework to achieve this.

Statistical uncertainties are the result of stochastic fluctations arising from the fact that a measurement is based on a finite set of observations. Repeated measurements of the same phenomenon will therefore result in a set of observations that will differ, and the statistical uncertainty is a measure of the range of this variation. By definition, statistical variations between two identical measurements of the same phenomenon are uncorrelated, and we have well-developed theories of statistics that allow us to predict and take account of such uncertainties in measurement theory, in inference and in hypothesis testing (see, for example, [1]). Examples of statistical uncertainties include the finite resolution of an instrument, the Poisson fluctations associated with measurements involving finite sample sizes and random variations in the system one is examining.

Systematic uncertainties, on the other hand, arise from uncertainties associated with the nature of the measurement apparatus, assumptions made by the experimenter, or the model used to make inferences based on the observed data. Such uncertainties are generally correlated from one measurement to the next, and we have a limited and incomplete theoretical framework in which we can interpret and acccommodate these uncertainties in inference or hypothesis testing. Common examples of systematic uncertainty include uncertainties that arise from the calibration of the measurement device, the probability of detection of a given type of interaction (often called the "acceptance" of the detector), and parameters of the model used to make inferences that themselves are not precisely known. The definition of such uncertainties is often ad hoc in a given measurement, and there are few broadly-accepted techniques to incorporate them into the process of statistical inference.

All that being said, there has been significant thought given to the practical problem of how to incorporate systematic uncertainties into a measurement. Examples of this work include proposals to combine statistical and systematic uncertainties into setting confidence limits on measurements [2, 3, 5, 6], techniques to estimate the magnitude of systematic uncertainties [4], and the use of standard statistical techniques to take into account systematic uncertainties [7, 8]. In addition, there have been numerous papers published on the systematic uncertainties associated with a given measurement [9].

In this review, I will first discuss a few case studies that illustrate how systematic uncertainties enter into some current measurements in high energy physics and astrophysics. I will then discuss a way in which one can consistently identify and characterize systematic uncertainties. Finally, I will outline the various techniques by which statistical and systematic uncertainties can be formally treated in measurements.

## 2. Case Studies

### 2.1. $W$ Boson Cross Section: Definitions are Relative

The production of the charged intermediate vector boson, the $W$, in proton-antiproton ($p\bar{p}$) annihilations is predicted by the Standard Model, and the measurement of its rate is of interest in high energy physics. This measurement involves counting the number of candidate events in a sample of observed interactions, $N_c$, estimating the number of "background" events in this sample from other processes, $N_b$, estimating the acceptance of the apparatus including all selection requirements used to define the sample of events, $\epsilon$, and counting the number of $p\bar{p}$ annihiliations, $L$. The cross section for $W$ boson production is then

$$\sigma_W = \frac{N_c - B_b}{\epsilon L}. \tag{1}$$

The CDF Collaboration at Fermilab has recently performed such a measurement [10], as illustrated in Fig. 1 where the transverse mass of a sample of candidate $W \to e\nu_e$ decays is illustrated. The measurement is quoted as

$$\sigma_W = 2.64 \pm 0.01(\mathrm{stat}) \pm 0.18(\mathrm{syst})\ \mathrm{nb}, \tag{2}$$

where the first uncertainty reflects the statistical uncertainty arising from the size of the candidate sample (approximately 38,000 candidates) and the second uncertainty arises from the background subtraction in Eq. (1). We can estimate these uncertainties as

$$\sigma_{stat} = \sigma_0/\sqrt{N_c} \tag{3}$$

$$\sigma_{syst} = \sigma_0\sqrt{\left(\frac{\delta N_b}{N_b}\right)^2 + \left(\frac{\delta\epsilon}{\epsilon}\right)^2 + \left(\frac{\delta L}{L}\right)^2}, \tag{4}$$

where the three terms in $\sigma_{syst}$ are the uncertainties arising from the background estimate $\delta N_b$, the acceptance $\delta\epsilon$ and the integrated luminosity $\delta L$. The parameter $\sigma_0$ is the measured value.

In the same sample, the experimenters also observe the production of the neutral intermediate vector boson, the $Z$. Because of this, the experimenters can measure the acceptance $\epsilon$ by taking a sample of $Z$ bosons identified by the two charged electrons they decay into, and then measuring $\epsilon$ from this sample. The dominant undertainty in this measurement arises from the finite statistics in the $Z$ boson sample. Thus, one could equivalently consider $\delta\epsilon$ to be a statistical uncertainty (and not a systematic one). This means that the uncertainties could have just as well been defined as

$$\sigma_{stat} = \sigma_0\sqrt{1/N_c + \left(\frac{\delta\epsilon}{\epsilon}\right)^2} \tag{5}$$

$$\sigma_{syst} = \sigma_0\sqrt{\left(\frac{\delta N_b}{N_b}\right)^2 + \left(\frac{\delta L}{L}\right)^2}. \tag{6}$$

resulting in a different assignment of statistical and systematic uncertainties.

Why would this matter? If we return back to our original discussion of what defines a statistical and systematic uncertainty, we normally assume a systematic uncertainty is correlated with subsequent measurements and it does not scale with the sample size.

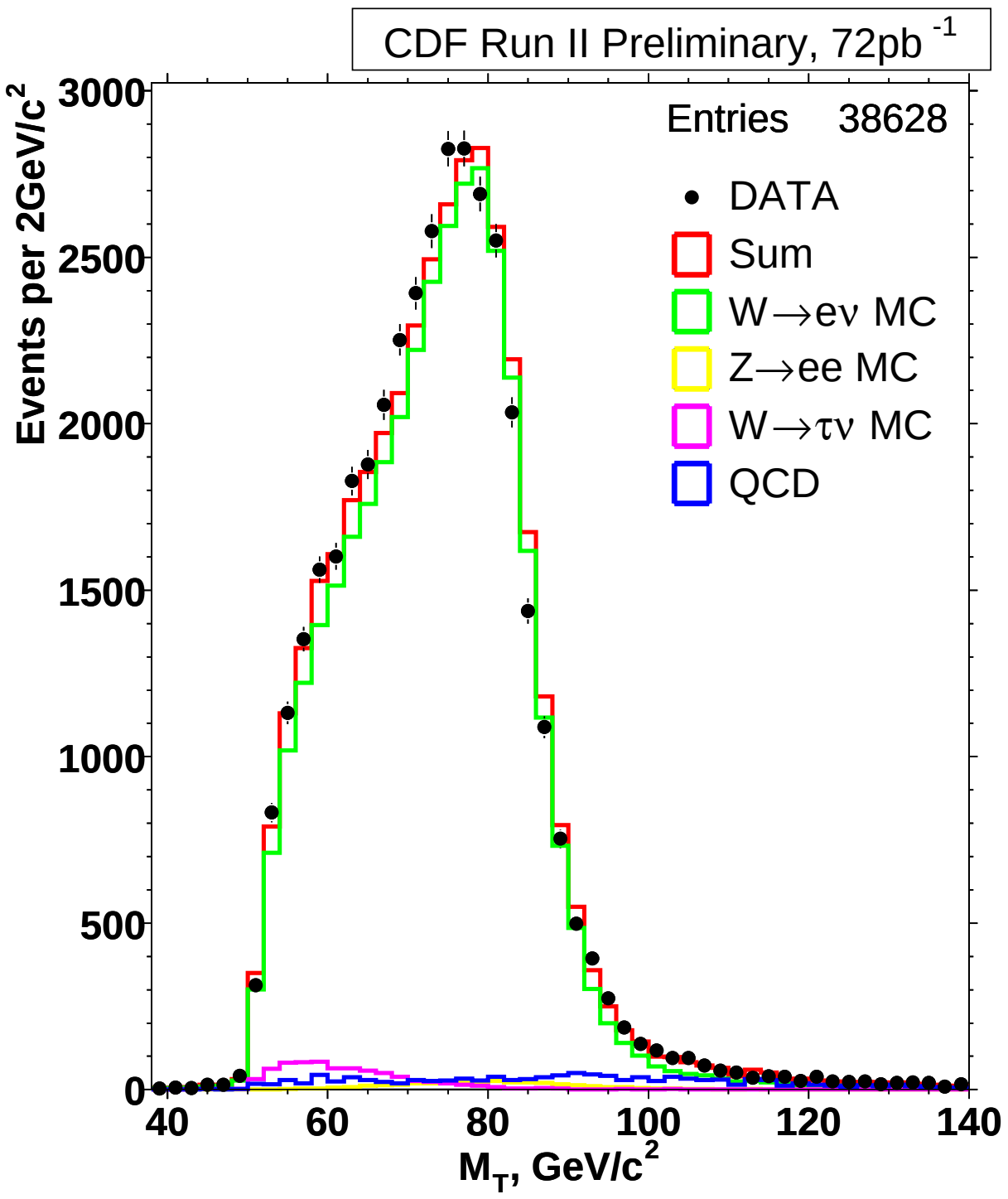

Figure 1: The transverse mass distribution for the $W$ boson candidates as observed recently by CDF. The peak reflects the Jacobian distribution typical of $W$ boson decays. The points are the measured distribution and the various histograms are the predicted distribution from $W$ decays and the other background processes.

In this case, the uncertainty on $\epsilon$ does not meet these requirements. The acceptance is a stochastic variable, which will become better known with increasing $Z$ boson sample size. It is therefore more informative to identify it as a statistical uncertainty. I will call this a "class 1" systematic uncertainty. Note that it would be appropriate to include in this category those systematic uncertainties that are in fact constrained by the result of a separate measurement, so long as the resulting uncertainty is dominated by the stochastic fluctuations in the measurement. An example of this could be the calibration constants for a detector that are defined by separate measurements using a calibration procedure whose precision is limited by statistics.

## 2.2. Background Uncertainty

The second case study also involves the measurement of $\sigma_W$ introduced in the previous section. The estimate of the uncertainty on the background rate $\delta N_b$ is performed by evaluating the magnitude of the different sources of candidate events that satisfy the criteria used to define the $W$ boson candidate sample. In the CDF measurement, it turns out that the background is dominated by events that arise from the pro-duction of two high-energy quarks or gluons (so-called "QCD events"), one of which "fakes" an electron or muon. A reliable estimate of this background is difficult to make from first principles, as the rate of such QCD events is many orders of magnitude larger than the $W$ boson cross section, and the rejection power of the selection criteria is difficult to measure directly.

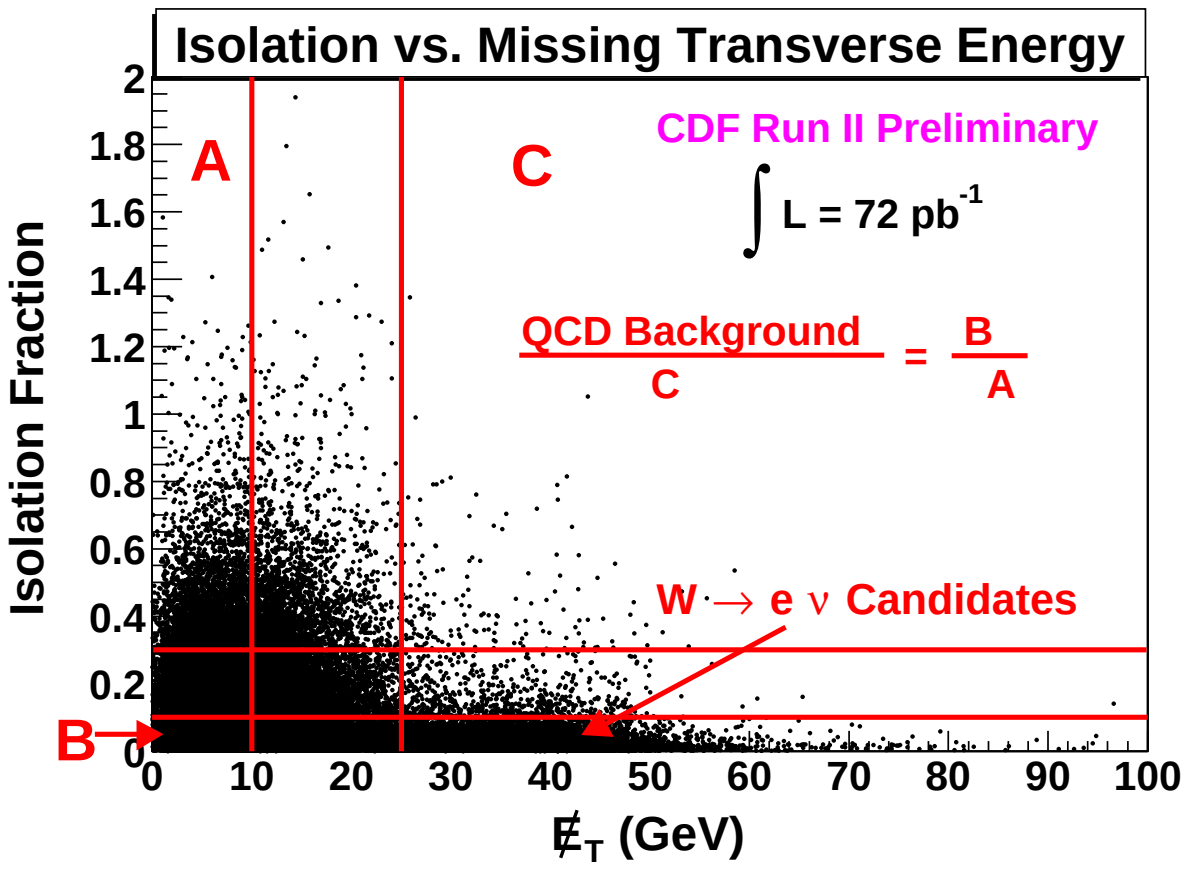

Figure 2: The distribution of the isolation fraction of the lepton candidate versus the missing energy in the event. Candidate leptons are required to have low isolation fractions ($< 0.10$), and QCD background events dominate the region with low missing transverse energy. The signal sample is defined by the requirement $\not{E}_T > 25$ GeV. The QCD background in the sample is estimated by the formula in the figure, which assumes that the isolation properties of the QCD events are uncorrelated with the missing transverse energy.

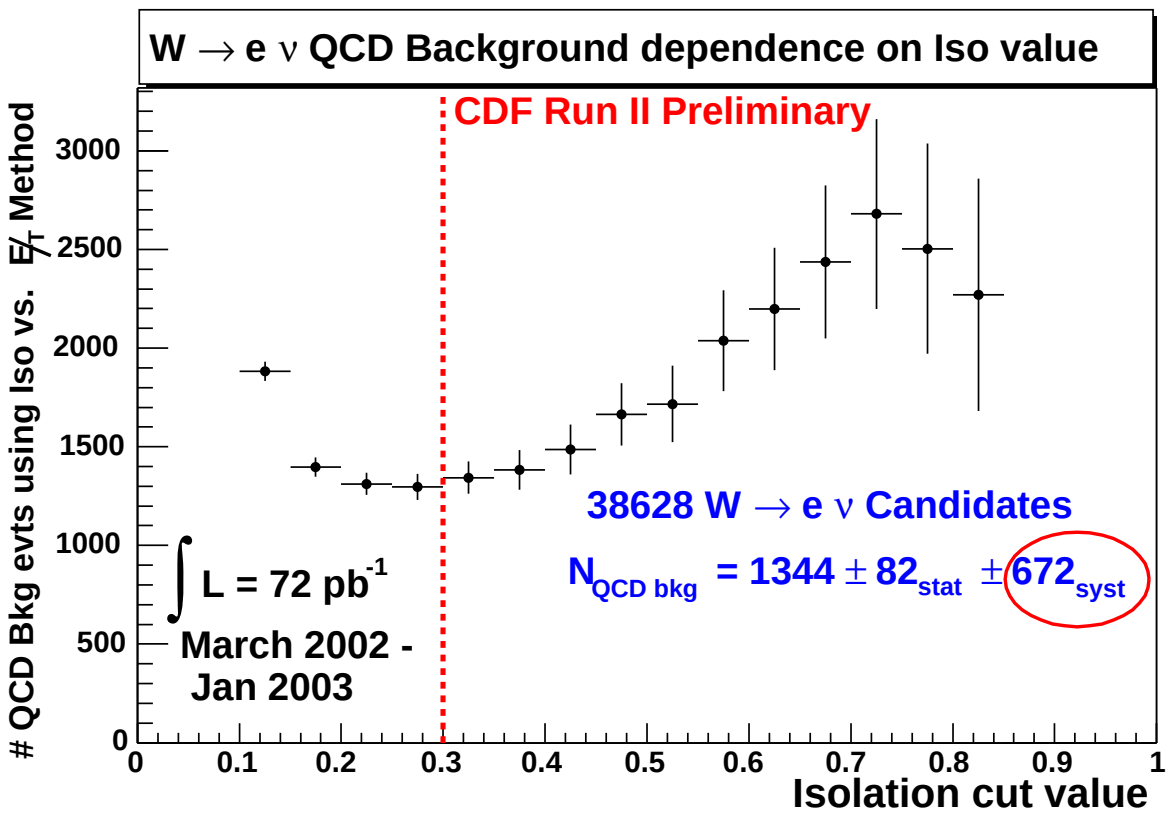

Figure 3: The variation of the background estimate as the isolation cut value is varied from 0.3 (the default value) up to 0.8. The isolation variable is a measure of the fraction of energy observed in a cone near the electron candidate normalized to the energy of the electron candidate.

The technique used to estimate $N_b$ in the CDF analysis is to take advantage of a known correlation: candidate events from QCD background will have more particles produced in proximity to the electron or muon candidate. At the same time, most of the QCD events will also have small values of missing transverse energy ($\not{E}_T$) compared with the $W$ boson events where a high-energy neutrino escapes undetected. Thus, a measure of the isolation of the candidate lepton and the $\not{E}_T$ can be an instrument to extract an estimate of the background in the observed sample. This is shown in Fig. 2, where one sees in the bottom-right region the signal region for this analysis. The region with low missing transverse energy is populated by QCD background events.

The dominant uncertainty in the background calculation arises from the assumption that the isolation properties of the electron candidate in QCD events is uncorrelated with the missing transverse energy in the event. Any such correlation is expected to be very small, and this is consistent with other observations. However, even a small correlation in these two variables results in a bias in the estimate of $N_b$. This potential bias is difficult to estimate with any precision. In this case, the experimenters varied the choice of the isolation criteria to define the QCD background region and the signal region, and used the variation in the background estimate as a measure of the systematic uncertainty $\delta N_b$. This variation is shown in Fig. 3.

This is an illustration of a systematic uncertainty that arises from one's limited knowledge of some features of the data that cannot be constrained by observations. In these cases, one often is forced to make some assumptions or approximations in the measurement procedure itself that have not been verified precisely. The magnitude of the systematic uncertainty is also difficult to estimate as it is not well-constrained by other measurements. In this sense, it differs from the class 1 systematic uncertainties introduced above.

I will therefore call this a "class 2" systematic uncertainty. It is one of the most common categories of systematic uncertainty in measurements in astrophysics and high energy physics.

## 2.3. Boomerang CMB Analysis

My third case study involves the analysis of the data collected by the Boomerang cosmic microwave background (CMB) probe, which mapped the spatial anisotropy of the CMB radiation over a large portion of the southern sky [11]. The data itself is a fine-grained two-dimensional plot of the spatial variation the temperature of part of the southern sky, as illustrated in Fig. 4. The analysis of this data involves the transformation of the observed spatial variation into a power series in spherical harmonics, with the spatial variations now summarized in the power spectrum as a function of the order of the spherical harmonic. The power spectrum includes all sources of uncertainty, including instrumental effects and uncertainties in cali-

brations.[1]

The Boomerang collaborators then test a large class of theoretical models of early universe development by determining the power spectrum predicted by each model and comparing the predicted and observed power as a function of spherical harmonic. These models are described by a set of cosmological parameters, each of them being constrained by other observations and theoretical assumptions. To determine those models that best describe the data, the experimenters take a Bayesian approach [12], creating a six-dimensional grid consisting of 6.4 million points, and calculating the likelihood function for the data at each point. They then define priors for each of the six parameters, and define a posterior probability that is now a function of these parameters. To make inferences on such key parameters as the age of the universe or its overall energy density, a marginalization is performed by numerically integrating the posterior probability over the other parameters. The experimenters can also consider the effect of varying the priors to explore the sensitivity of their conclusions to the priors themselves.

In this analysis, the lack of knowledge in the paradigm used to make inferences from the data is captured in the choice of priors for each of the parameters. A classical statistical approach could have equivalently defined these as sources of systematic uncertainty. Viewed from either perspective, the uncertainties that arise from the choice of paradigm are not statistical in nature, given that they would affect any analysis of similar data. Yet they differ from the two previous classes of systematic uncertainty I have identified, which arise directly from the measurement technique. I therefore define such theoretically-motivated uncertainties as "class 3" systematics. I also note that the Bayesian technique to incorporate these uncertainties has no well-defined frequentist analogue, in that one cannot readily identify an ensemble of experiments that would replicate the variation associated with these uncertainties.

The distinction between class 2 and class 3 systematics comes in part from the fact that one is associated with the measurement technique while the other arises in the interpretation of the observations. I argue, however, that there is an additional difference: In the first case, there is a specific piece of information needed to complete the measurement, the background yield $N_b$, and the systematic uncertainty arises from manner in which that is estimated. In the other case, the experiment measures the spatial variation in the CMB and summarizes these data in the multipole moments. The systematic uncertainties that are associated with the subsequent analysis of these data in terms of cosmological parameters is very model-dependent, and the systematic uncertainties arise from the attempt to extract information about a subset of the parameters in the theory (for example, the age of the universe or the energy density).

[1]The uncertainties associated with instrumental effects and calibrations are also systematic in nature, but we will not focus on these here.

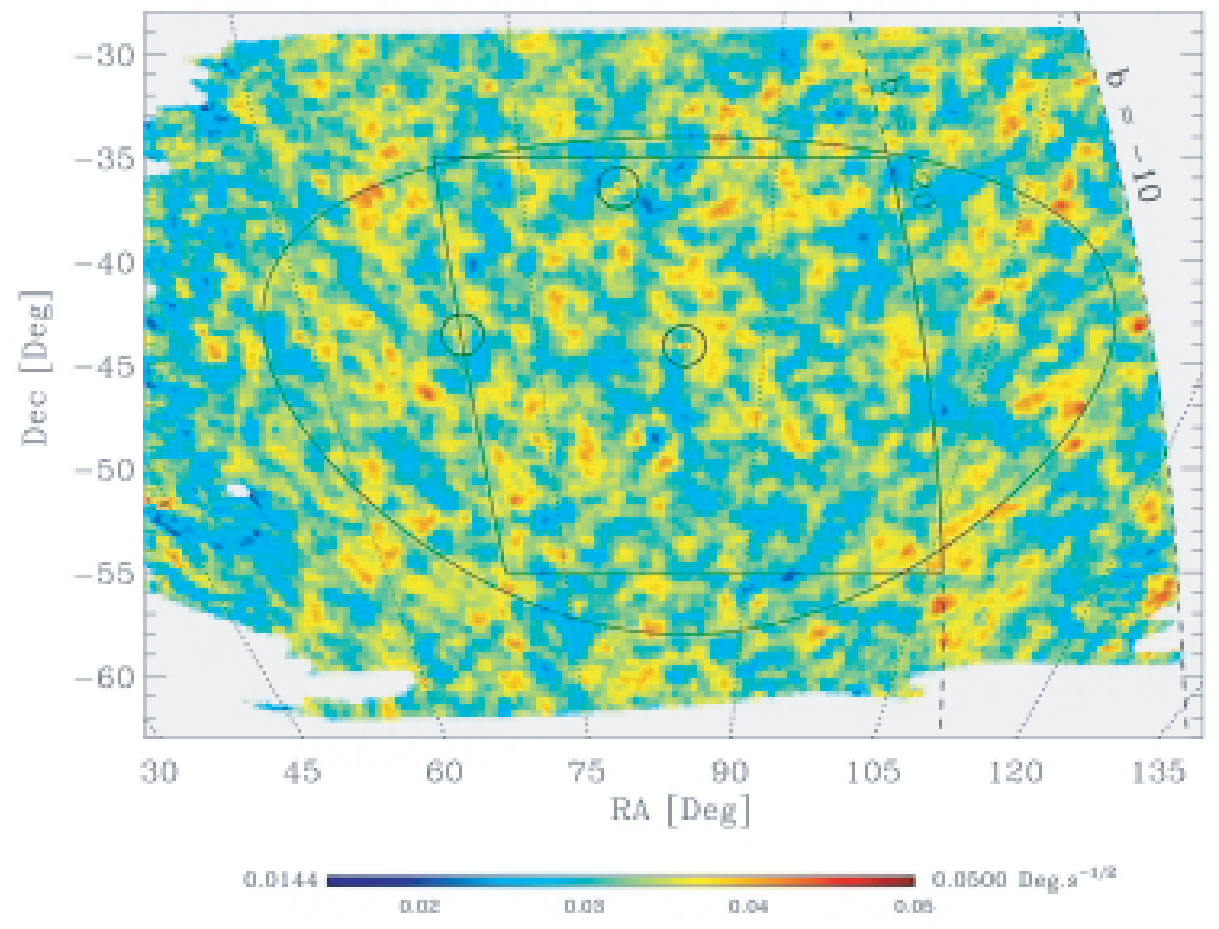

Figure 4: The temperature variation of the CMB as measured by the Boomerang experiment. The axes represent the declination and azimuth of the sky, and the contour is the region used in the analysis.

## 2.4. Summary of Taxonomy

In these case studies, I have motivated three classes of systematic uncertainties. Class 1 systematics are uncertainties that can be constrained by ancillary measurements and can therefore be treated as statistical uncertainties. Class 2 systematics arise from model assumptions in the measurement or from poorly understood features of the data or analysis technique that introduce a potential bias in the experimental outcome. Class 3 systematics arise from uncertainties in the underlying theoretical paradigm used to make inferences using the data.

The advantages of this taxonomy are several. Class 1 systematics are statistical in nature and will therefore naturally scale with the sample size. I recommend that they be properly considered a statistical uncertainty and quoted in that manner. They are not correlated with independent measurements and are therefore straightforward to handle when combining measurements or making inferences. Class 2 systematics are the more challenging category, as they genuinely reflect some lack of knowledge or uncertainty in the model used to analyze the data. Because of this, they also have correlations that should be understood in any attempt to combine the measurement with other observations. They also do not scale with the sample size, and therefore may be fundamental limits on how well one can perform the measurement. Class 3 systematics do not depend on how well we understand the measurement per se, but are fundamentally tied to the theoretical model or hypothesis being tested. As such, there is significant variation in practice. As is illustrated in the third case study and as I will discuss below, a Bayesian approach allows for a range of possible models to be tested if one can parametrize the uncertainties in the relevant probability distribution function and then define reasonable priors. A purely frequentist approach to this problem founders on how one would define the relevant ensemble.

## 3. Estimation of Systematic Uncertainties

There is little, if any, formal guidance in the literature for how to define systematic uncertainties or estimate their magnitudes, and much of current practice has been defined by informal convention and "oral tradition." A fundamental principle, however, is that the technique used to define and estimate a systematic uncertainty should be consistent with how the statistical uncertainties are defined in a given measurement, since the two sources of uncertainty are often combined in some way when the measurement is compared with theoretical predictions or independent measurements.

Perhaps the most challenging aspect of estimating systematic uncertainties is to define in a consistent manner all the relevant sources of systematic uncertainty. This requires a comprehensive understanding of the nature of the measurement, the assumptions implicit or explicit in the measurement process, and the uncertainties and assumptions used in any theoretical models used to interpret the data. In any robust design of an experiment, the experimenters will anticipate all sources of systematic uncertainty and should design the measurement to minimize or constrain them appropriately. Good practice suggests that the analysis of systematic uncertainties should be based on clear hypotheses or models with well-defined assumptions.

In the process of the measurement, it is often typical to make various "cross-checks" and tests to determine that no unanticipated source of systematic uncertainty has crept into the measurement. A cross-check, however, should not be construed as a source of systematic uncertainty (see, for example the discussion in [4]).

A common technique for estimating the magnitude of systematic uncertainties is to determine the maximum variation in the measurement, $\Delta$, associated with the given source of systematic uncertainty. Arguments are then made to transform that into a measure that corresponds to a one standard deviation measure that one would associate with a Gaussian statistic, with typical conversions being $\Delta/2$ and $\Delta/\sqrt{12}$, the former being argued as a deliberate overestimate, and the latter being motivated by the assumption that the actual bias arising from the systematic uncertainty could be anywhere within the interval $\Delta$. Since it is common in astrophysics and high energy physics to quote 68% confidence level intervals as statistical uncertainties, it therefore is appropriate to estimate systematic uncertainties in a comparable manner.

There are various practices that tend to overestimate the magnitude of systematic uncertainties, and these should be avoided if one is to not dilute the statistical power of the measurement. A common mistake is to estimate the magnitude of a systematic uncertainty by using a shift in the measured quantity when some assumption is varied in the analysis technique by what is considered the relevant one standard deviation interval. The problem with this approach is that often the variation that is observed is dominated by the statistical uncertainty in the measurement, and any potential systematic bias is therefore obscured. In such cases, I recommend that either a more accurate procedure be found to estimate the systematic uncertainty, or at the very least that one recognize that this estimate is unreliable and likely to be an overestimate. A second common mistake is to introduce a systematic uncertainty into the measurement without an underlying hypothesis to justify the concern. This is often the result of confusing a source of systematic uncertainty with a "cross check" of the measurement.

## 4. The Statistics of Systematic Uncertainties

A reasonable goal in any treatment of systematic uncertainties is that consistent and well-established procedures be used that allow one to understand how to best use the information embedded in the systematic uncertainty when interpreting the measurement. Increasingly, the fields of astrophysics and high energy physics have developed more sophisticated approaches to interval estimation and hypothesis testing. Frequentist approaches have returned to the fundamentals of Neyman constructions and the resulting coverage properties. Bayesian approaches have explored the implications of both objective and subjective priors, the nature of inference and the intrinsic power embedded in such approaches when combining information from multiple measurements.

I will outline how systematic uncertainties can be accommodated formally in both Bayesian and frequentist approaches.

## 4.1. Formal Statement

To formally state the problem, assume we have a set of observations $x_i, i = 1, n$, with an associated probability distribution function $p(x_i|\theta)$, where $\theta$ is an unknown random parameter. Typically, we wish to make inferences about $\theta$. Let us now assume that there is some additional uncertainty in the probability distribution function that can be described with another unknown parameter $\lambda$. This allows us to define a likelihood function

$$\mathcal{L}(\theta, \lambda) = \prod_i p(x_i|\theta, \lambda). \quad (7)$$

Formally, one can treat $\lambda$ as a "nuisance parameter." In many cases (especially those associated with class 1 systematics), one can identify a set of additional observations of a random statistic $y_j, j = 1, m$ that provides information about $\lambda$. In that case, the likelihood would become

$$\mathcal{L}(\theta, \lambda) = \prod_{i,j} p(x_i, y_j|\theta, \lambda). \quad (8)$$

With this formulation, one sees that one has to find a means of taking into account the uncertainties that arise from the presence of $\lambda$ in order to make inferences on $\theta$. I will discuss some possible approaches.

## 4.2. Bayesian Approach

A Bayesian approach would involve identification of a prior, $\pi(\lambda)$, that characterizes our knowledge of $\lambda$. Typical practice has been to either assume a flat prior or, in cases where there are corollary measurements that give us information on $\lambda$, a Gaussian distribution. One can then define a Bayesian posterior probability distribution

$$\mathcal{L}(\theta, \lambda)\ \pi(\lambda)\, d\theta d\lambda, \quad (9)$$

which we can then marginalize to set Bayesian credibility intervals on $\theta$.

This is a straightforward statistical approach and results in interval estimates that can readily be interpreted in the Bayesian context. The usual issues regarding the choice of priors remains, as does the interpretation of a Bayesian credibility interval. These are beyond the scope of this discussion, but are covered in most reviews of this approach [12].

## 4.3. Frequentist Approach

The frequentist approach to the formal problem also starts with the joint probability distribution $p(x_i, y_j|\theta, \lambda)$. There are various techniques for how to deal with the presence of the nuisance parameter $\lambda$, and I will outline just a few of them. I will note that there isn't a single commonly adopted strategy in the literature, and even the simplest techniques tend to involve significant computational burden.

One technique involves identifying a transformation of the parameters to factorize the problem in such a manner that one can then integrate out one of the two parameters [13]. This approach is robust and theoretically sound, and in the trivial cases results in a 1-dimensional likelihood function that now incorporates the uncertainties arising from the nuisance parameter. It has well-defined coverage properties and a clear frequentist interpretation. However, this approach is of limited value given that it is necessary to find an appropriate transformation.

I note that this approach is only of value in cases where one is dealing with a class 1 systematic uncertainty that is, as I have argued above, formally a source of statistical uncertainty. Class 2 and class 3 systematic uncertainties cannot be readily constrained by a set of observations represented by the $y_j, j = 1, m$.

A second approach to the incorporation of nuisance parameters is to define Neyman "volumes" in the multi-dimensional parameter space, equivalent to what is done in the case of a interval setting with one random parameter. In this case, one creates an infinite set of two dimensional contours defined by requiring that the observed values lie within the contour the necessary fraction of the time (say 68%). Then one identifies the locus of points in this two-dimensional space defined by the centres of each contours, and this boundary becomes the multi-dimensional Neyman interval for both parameters, as illustrated in Fig. 5. To "eliminate" the nuisance parameter, one projects the two-dimensional contour onto the axis of the parameter of interest. This procedure results in a frequentist confidence interval that over-covers, and in some cases over-covers badly. It thus results in "conservative" intervals that may diminish the statistical power of the measurement.

A third technique to take into account systematic uncertainties involves what is commonly called the "profile method" where one eliminates the nuisance parameter by creating a profile likelihood defined as the value of the likelihood maximized by varying $\lambda$ for each value of the parameter $\theta$ [14]. This creates a likelihood function independent of $\lambda$, but one that has ill-defined coverage properties that depend on the correlation between $\lambda$ and $\theta$. However, it is a straightforward technique, that is used frequently and that results in inferences that are at some level conservative.

A variation on these techniques has been used in recent analyses of solar neutrino data, where the analysis uses 81 observables and characterizes the various systematic uncertainties by the introduction of 31 parameters [15]. They linearize the effects of the systematic parameters on the chi-squared function and then minimize the chi-squared with respect to each of the parameters. In this sense, this analysis is a concrete example of the profile method.

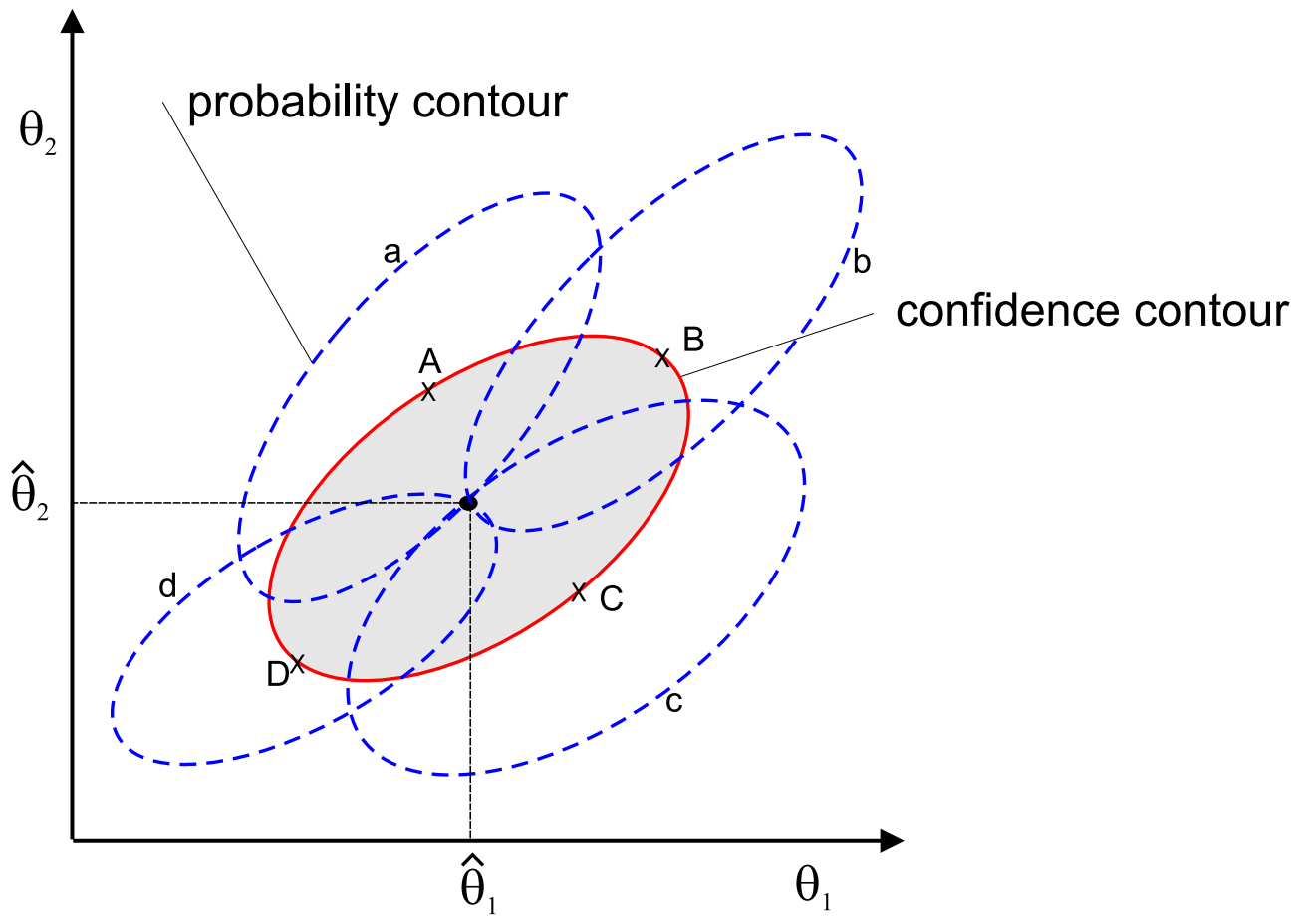

Figure 5: The Neyman construction for two parameters. The nuisance parameter is $\theta_2$, and the shaded region is the interval defined by this construction. The four dashed contours are examples of the intervals that define the contour (from G. Zech).

Although most of these techniques are approximate in some sense, they have the virtue that they scale in an intuitively acceptable manner. In the limit where the statistical uncertainties are far larger than the systematic uncertainties, the former are the dominant effect in any inference or hypothesis test. Conversely, when the systematic uncertainties begin to compete with or dominate the statistical uncertainties, the results of any statistical inference reflect the systematic uncertainties and these become the limiting factor in extracting information from the measurement. I would argue that this should be a minimal requirement of any procedure used to take into account systematic uncertainties.

## 4.4. Hybrid Techniques

There has been one technique in common use in high energy physics to incorporate sources of systematic uncertainty into an analysis, first described by R. Cousins and V. Highland [2]. Using the notation introduced earlier, the authors argue that one should create a modified probability distribution function

$$p_{CH}(x|\theta) = \int p(x|\theta,\lambda)\pi(\lambda)\, d\lambda, \qquad (10)$$

which could be used to define a likelihood function and make inferences on $\theta$. They argue that this can be understood as approximating the effects of having an ensemble of experiments each of them with various choices of the parameter $\lambda$ and with the distribution $\pi(\lambda)$ representing the frequency distribution of $\lambda$ in this ensemble.

Although intuitively appealing to a physicist, this approach does not correspond to either a truly frequentist or Bayesian technique. On the one hand, the concept of an ensemble is a frequentist construct. On the other hand, the concept of integrating or "averaging" over the probability distribution function is a Bayesian approach. Because of this latter step, it is difficult to define the coverage of this process [5]. I therefore consider it a Bayesian technique that can be readily understood in that formulation if one treats the frequency distribution $\pi(\lambda)$ as the prior for $\lambda$. I note that it also has the desired property of scaling correctly as one varies the relative sizes of the statistical and systematic uncertainties.

## 5. Summary and Conclusions

The identification and treatment of systematic uncertainties is becoming increasingly "systematic" in high energy physics and astrophysics. In both fields, there is a recognition of the importance of systematic uncertainties in a given measurement, and techniques have been adopted that result in systematic uncertainties that can be compared in some physically relevant sense with the statistical uncertainties.

I have proposed that systematic uncertainties can be classified into three broad categories, and by doing so creating more clarity and consistency in their treatment from one measurement to the next. Such classification, done a priori when the experiment is being defined, will assist in optimizing the experimental design and introducing into the data analysis the necessary approaches to control and minimize the effect of these systematic effects. In particular, one should not confuse systematic uncertainties with cross-checks of the results.

Bayesian statistics naturally allow us to incorporate systematic uncertainties into the statistical analysis by introducing priors for each of the parameters associated with the sources of systematic uncertainty. However, one must be careful regarding the choice of prior. I recommend that in all cases the sensitivity of any inference or hypothesis test to the choice of prior be investigated in order to ensure that the conclusions are robust.

Frequentist approaches to systematic uncertainties are less well-understood. The fundamental problem is how one defines the concept of an ensemble of measurements, when in fact what is varying is not an outcome of a measurement but ones assumptions concerning the measurement process or the underlying theory. I am not aware of a robust method of incorporating systematic uncertainties in a frequentist paradigm except in cases where the systematic uncertainty is really a statistical uncertainty and the additional variable can be treated as a nuisance parameter. However, the procedures commonly used to incorporate systematic uncertainties into frequentist statistical inference do have some of the desired "scaling" properties.

## Acknowledgments

I gratefully acknowledge the support of the Natural Sciences and Engineering Research Council of Canada, and thank the conference organizers for their gracious hospitality.

## References

[1] A. Stuart and J. K. Ord, *Kendall's Advanced Theory of Statistics,* Oxford University Press, New York (1991).

[2] R. D. Cousins and V. L. Highland, Nucl. Instrum. Meth. **A320**, 331 (1992).

[3] C. Guinti, Phys. Rev. D **59**, 113009 (1999).

[4] R. J. Barlow, "Systematic Errors, Fact and Fiction," hep-ex/0207026 (Jun 2002). Published in the Proceedings of the Conference on Advanced Statistical Techniques in Particle Physics, Durham England, Mar 16-22, 2002. See also, R. J. Barlow, "Asymmetric Systematic Errors," hep-ph/0306138 (Jun 2003).

[5] J. Conrad *et al.*, Phys. Rev. D **67**, 012002 (2003).

[6] G. C. Hill, Phys. Rev. D **67**, 118101 (2003).

[7] G. Zech, Eur. Phys. J **C4:12** (2002).

[8] L. Demortier, " Bayesian Treatment of Systematic Uncertainties," Published in the Proceedings of the Conference on Advanced Statistical Techniques in Particle Physics, Durham England, Mar 16-22, 2002.

[9] See, for example: A. G. Kim *et al.*, "Effects of Systematic Uncertainties on the Determination of Cosmological Parameters," astro-ph/0304509 (Apr 2003).

[10] T. Dorigo *et al.* (The CDF Collaboration), FERMILAB-CONF-03/193-E. Published in the Proceedings of the 38th Rencontres de Moriond on QCD and High-Energy Hadronic Interactions, Les Arcs, Savoie, France, March 22-29, 2003.

[11] B. Netterfield *et al.*, Ap. J. **571**, 604 (2002).

[12] E. T. Jaynes, *Probability Theory: The Logic of Science*, Cambridge University Press (1995).

[13] G. Punzi, "Including Systematic Uncertainties in Confidence Limits," CDF Note in preparation (Jul 2003).

[14] See, for example, N. Reid, in these proceedings.

[15] Fogli *et al.*, Phys. Rev. D **66**, 053010 (2002).